\documentclass[preprintnumbers,showpacs, showkeys, twocolumn,secnumarabic,amssymb,amsmath,amsfonts,aps, pre]{revtex4-1}
\usepackage{graphicx}
\usepackage{comment}
\usepackage{amsmath}
\usepackage{xcolor}
\usepackage{subfigure}
\usepackage{color}

\begin{document}

\newcommand{\C}{\textbf C}
\newcommand{\N}{\mathbb N}
\newcommand{\Z}{\mathbb Z}
\newcommand{\R}{\mathbb R}
\newcommand{\A}{\mathbf A}
\newcommand{\K}{\mathbf K}
\newcommand{\I}{\mathbf I}

\newtheorem{theorem}{Theorem}[section]
\newtheorem{lemma}[theorem]{Lemma}
\newtheorem{proposition}[theorem]{Proposition}
\newtheorem{corollary}[theorem]{Corollary}
\newtheorem{remark}[theorem]{Remark}
\newtheorem{definition}[theorem]{Definition}
\newtheorem{example}[theorem]{Example}

\newcommand{\SG}{\mathcal G}
\newcommand{\CC}{\mathrm{C}}
\newcommand{\PC}{\mathcal{P}}
\newcommand{\T}{\mathcal{T}_f}
\newcommand{\Tz}{\mathcal{T}_f^0}

\newcommand{\m}{\mathfrak{M}}
\newcommand{\di}{\mathrm{d}}
\newcommand{\End}{\mathrm{End}}

\newcommand{\Mat}{\mathrm{Mat}}
\newcommand{\GL}{\mathrm{GL}}
\newcommand{\PP}{\mathrm{P}}
\newcommand{\zz}{\mathbb{Z}}
\newcommand{\NN}{\mathbb{N}}
\newcommand{\HH}{\mathrm{H}}
\newcommand{\HHS}{\mathrm{H}^\mathrm{S}}
\newcommand{\ZZ}{\mathrm{Z}}
\newcommand{\BB}{\mathrm{B}}

\newcommand{\Mod}{\mathop{\mathrm Mod}}

\newcommand{\al}{\alpha}
\newcommand{\bt}{\beta}
\newcommand{\om}{\Omega}
\newcommand{\mm}{\omega}
\newcommand{\ups}{\upsilon}
\newcommand{\up}{\Upsilon}
\newcommand{\gm}{ k_1}

\newcommand{\B}{\mathcal{B}}
\newcommand{\D}{\mathcal{D}}
\newcommand{\M}{\mathcal{M}}
\newcommand{\h}{\mathcal{H}}
\newcommand{\Y}{\mathcal{Y}}

\newcommand{\pp}{\mathcal{P}}
\newcommand{\nn}{\mathcal{N}}

\newcommand{\G}{\mathcal{G}}
\newcommand{\Ord}{\mathcal{O}}

\newcommand{\KK}{\mathfrak{K}}

\newcommand{\gr}{\mathfrak{G}}

\newcommand{\ff}{\mathfrak{F}}
\newcommand{\abs}[1]{\lvert#1\rvert}

\newcommand{\aaa}{\mathcal{A}}
\newcommand{\sss}{\mathcal{S}}
\newcommand{\cl}{\mathrm{Cl}}

\everymath{\displaystyle}
%%%%%%%%%%%%%%%%%%%%%%%%%%%%%%%%%%%%%%%%%%%%%%%%%%%%%%%%%
% AUTHOR INFORMATIONS
\title{Persistent Homology analysis of  Phase Transitions}
\date{\today}
\author{Irene Donato}
\email{irene.irened@gmail.com}
\author{Matteo Gori}
\email{gori6matteo@gmail.com}
\author{Marco Pettini}
\email{pettini@cpt.univ-mrs.fr}
\affiliation{Aix-Marseille University, CNRS Centre de Physique Th\'eorique UMR 7332,
Campus de Luminy, Case 907, 13288 Marseille Cedex 09, France}
\author{Giovanni Petri}
\email{giovanni.petri@isi.it}
\affiliation{ ISI Foundation, Turin, Italy}
\author{Sarah De Nigris}
\email{denigris.sarah@gmail.com}
\affiliation{NaXys, D\'{e}partement de Math\'{e}matique, Universit\'{e} de Namur,
8 repart de la Vierge, 5000 Namur, Belgique}
\author{Roberto Franzosi}
\email{bob.franzosi@gmail.com}
\affiliation{Qstar, Istituto Nazionale di Ottica, largo E. Fermi 6, 50125 Firenze, Italy}
\author{Francesco Vaccarino}
\email{francesco.vaccarino@gmail.com}
\affiliation{Dipartimento di Scienze Matematiche "G.L.Lagrange", Politecnico di Torino, C.so Duca degli Abruzzi 24, and ISI Foundation, Turin, Italy}

%%%%%%%%%%%%%%%%%%%%%%%%%%%%%%%%%%%%%%%%%%%%%%%%%%%%%%%%
% ABSTRACT AND PACS
\begin{abstract}
\textit{Persistent homology} analysis, a recently developed computational method in algebraic topology, is applied to the study of the phase transitions  undergone by the so-called XY-mean field model 
and by the $\phi^4$ lattice model, respectively.
For both models the relationship between phase transitions and the topological properties of certain submanifolds of configuration space are exactly known.
It turns out that these a-priori known facts are clearly retrieved by persistent homology analysis of dynamically sampled submanifolds of configuration space.
\end{abstract}
%%%%%%%%%%%%%%%%%%%%%%%%%%%%%%%%%%%%%%%%%%%%%%%%%%%%%%%%%
\maketitle
%%%%%%%%%%%%%%%%%%%%%%%%%%%%%%%%%%%%%%%%%%%%%%%%%%%%%%%%%
% BODY OF PAPER
%%%%%%%%%%%%%%%%%%%%%%%%%%%%%%%%%%%%%%%%%%%%%%%%%%%%%%%%%
\section{Introduction. }
Topological methods lie at the base of many successful physical theories \cite{Nakahara}, with fields of applications ranging from dynamical systems and quantum computation to the theory of phase transitions and topological field theories. 
In recent years, it has been investigated \cite{PettiniBook} the possibility that at least for a broad class of physical systems the deep origin of phase transitions be a major topological change of some submanifolds of phase space or, equivalently, of configuration space. \\
The central idea is that the singular energy dependence displayed by the thermodynamic observables at a phase transition is the ”shadow” of such major topological change. 
This new point of view about 
the deep origin of phase transitions was originally proposed for theoretical reasons, in fact, after the Yang-Lee theorem the mathematical description of phase transitions requires the thermodynamic limit ($N\rightarrow\infty$) in order to break the analyticity of thermodynamic observables. However, phase transition phenomena occur in nature as dramatic qualitative changes of some physical property also very far from thermodynamic limit. Let us think of Bose-Einstein condensation, of Dicke's superradiance in "microlasers", of superconductive transitions in very small metallic objects, of the filament-globule transition in homopolymers, of the folding transition in proteins, of a microscopic snowflake melting into a droplet of liquid water, to mention just some examples. \\
The question was: can we think of a different mathematical approach unifying the description of phase transitions in finite, small $N$ systems with the standard description resorting to the thermodynamic limit dogma? At least for a broad class of physical potentials the answer was in the affirmative as can be seen in Refs.  \cite{PettiniBook}  and \cite{TH1,TH2}. 
However, in some sense similarly to the Yang-Lee theory for which analytically finding the zeros of the grand partition function is in practice possible only for a few models (essentially given by the Lee-Yang "circle theorem"), also the topological approach suffers of computational difficulties, and analytic
topological information can be obtained  only for a very few models.
Also the direct numerical measurement of topological properties of the configuration space of physical systems faces serious computational issues because of the high dimensionality of the associated manifolds.
The idea that some of the mentioned computational obstacles could be overcome comes from the observation of the existence of new computational tools in the fields of discrete geometry and topology. These new methods have already been developed for analysing data in high-dimensional spaces \cite{Niyogi:2006}. Hence, we expect that they could be useful to investigate topological changes also in physical configuration spaces by identifying their homology from random samples.\\
In the present paper  we resort to \textit{persistent homology} analysis. 
Persistent homology \cite{Ghrist:2008tw,Pers,Top}, a particular sampling-based technique from algebraic topology, was originally  introduced in 2002 \cite{edels:2002} by Edelsbrunner {\it et al} with the aim of extracting coarse topological information from high-dimensional datasets \cite{Niyogi:2006}. 
In a nutshell, while \textit{homology} detects the connected components, tunnels, voids of a given topological space, persistent homology computes multi-scale homological features obtained from a discrete sample of a topological space $X$ by foliating it appropriately.  
Hitherto, the study of persistent homology has already proved useful in various fields like biological and medical data analysis, neuroscience \cite{Petri:2014hq}, sensor network coverage problems \cite{DeSilva:2007ve}, to quote just a few of them.\\
Here persistent homology is applied to the study of equilibrium phase transitions. Two models are considered for which we rigorously know what to expect: the so-called Mean-Field XY model (MFXY) and the classical lattice $\phi^4$ model. 
For the MFXY model both the thermodynamics and the configuration space topology are exactly known, whence the topological origin 
of phase transition is rigorously ascertained; while for the $\phi^4$ model it is analytically known that the phase transition does not correspond to any topology change in configuration space at any \textit{finite} $N$ (see Section \ref{fi4} for a discussion on this model). \\
The benchmarking so performed gave sharp and unambiguous results in the good direction. 
This could  open new interesting perspectives for practical applications of the above mentioned topological theory of phase transitions. 
\section{Phase Transitions and Topology} 
Apart from several studies on specific models \cite{PettiniBook}, 
two theorems state that the unbounded growth with $N$ of certain thermodynamic quantities, eventually leading to singularities in the $N\to\infty$ limit - the hallmark
of an equilibrium phase transition - is {\it necessarily} due to appropriate
topological transitions in configuration space \cite{TH1,TH2}.  The following \textit{exact} formula 
\[
S_N(v) =({k_B}/{N}) \log \left[ \int_{M_v}\ d^Nq\right] 
\]
\[
=\frac{k_B}{N} \log \left[ vol
[{M_v \setminus\bigcup_{i=1}^{{\cal N}(v)} \Gamma(x^{(i)}_c)}]\ +
\sum_{i=0}^N w_i\ \mu_i (M_v)+ {\cal R} \right]  ,\label{exactS}
\]
makes \textit{explicit} the relation between thermodynamics and topology, 
where $S$ is the configurational entropy, $v$ is the potential energy per degree of freedom,  and the
$\mu_i(M_v)$ are the Morse indexes (in one-to-one correspondence
with topology changes) of the submanifolds $\{ M_{v}=V_N^{-1}((-\infty,v])\}_{v \in{\Bbb R}}$ of configuration
space; in square brackets: the first term is the result of the excision of certain neighborhoods of
the critical points of the interaction potential from  $M_{v}$; the second term
is a weighed sum of the Morse indexes, and the third term is a smooth function of $N$ and $v$.
It is evident that sharp changes in the potential energy pattern of at least some of
the $\mu_i(M_v )$ (thus of the way topology changes with $v$) affect $S(v)$ and its
derivatives. It can be proved that the occurrence of phase transitions necessarily
stems from this topological part of thermodynamic entropy \cite{TH1, TH2}.

\subsection{The Mean-Field $XY$ Model.}
\label{sec_MF}
The mean-field $XY$ model is defined by the Hamiltonian
\cite{Antoni:2005,review_longrange}
\begin{equation}
{\cal H}(p,\varphi )= \sum_{i=1}^N \frac{p_i^2}{2} +
 \frac{J}{2N}\sum_{i,j=1}^N
\left[ 1 - \cos(\varphi_i - \varphi_j)\right] -h\sum_{i=1}^N \cos\varphi_i ~.\nonumber
\label{Hmf}
\end{equation}
Here $\varphi_i \in [0,2\pi]$ is the rotation angle of the $i$th rotator
and $h$ is an external field. Defining at each site $i$ a
classical spin vector ${\bf m}_i = (\cos\varphi_i,\sin\varphi_i)$, the model
describes a planar ($XY$) Heisenberg system with interactions of equal strength
among all the spins. We consider the ferromagnetic case $J =1$.
The equilibrium statistical mechanics
of this system is exactly described, in the thermodynamic limit, by mean-field
theory. In the limit $h \to 0$, the system has a continuous
phase transition, with classical critical exponents, at the critical temperature $T_c = 1/2$, or
at the critical energy density $E_c/N = 3/4$ \cite{Antoni:2005}.

The entire configuration space $M$ of the model is an $N$-dimensional torus, 
parametrized by $N$ angles.
The submanifolds $M_v\subset M$ are defined by
\[
M_v = {\cal V}^{-1} (-\infty,v] 
\]
\vskip -0.8truecm
\begin{equation}
=\{ (\varphi_1,\dots,\varphi_N) \in M : {\cal V}(\varphi_1,\dots,\varphi_N) \leq v\}~,
\label{submanifolds}
\end{equation}
i.e., defined by the constraint that the potential energy per particle ${\cal V} = V/N$ does not exceed a given value $v$.

Morse theory \cite{milnor} states that topology changes of the $M_v$ occur in
correspondence with critical points of $\cal V$, i.e., those points where $\nabla {\cal V} =0$. This implies \cite{PettiniBook} that there are no topological changes for ${\cal V}> 1/2 + h^2/2$, i.e., all the $M_v$ with ${\cal V} > 1/2 + h^2/2$ are diffeomorphic to the whole $M$.
 
The Euler characteristic, a topological invariant of the manifolds $M_v$ which is {\it exactly} computed in Ref.\cite{xymf,physrep}, is defined by 
\begin{equation} 
\chi (M_v) = \sum_{k = 0}^N (-1)^k \mu_k(M_v)~, \label{chi_morse} 
\end{equation}
where the {\em Morse number} $\mu_k$ is the number of critical points of $\cal V$
that have index $k$  \cite{milnor}.

After a monotonic growth with $v$, a sharp, discontinuous jump to zero of $\chi (M_v)$ is found at the phase transition point, that is, at $v_c = {1}/{2} + 0^+$.
However, as already shown in \cite{xymf,physrep},
it is just this major topological change occurring at $v_c$  that is
related to the thermodynamic phase transition of the Mean Field $XY$ model.

\subsection{The $\phi^4$ model.}\label{fi4}
The lattice $\phi^4$ model is defined by the Hamiltonian 
\begin{equation}
{\cal H} ( p, \varphi )= \sum_i  \left[ \frac{p_i^2}{2} + \frac{J}{2} \sum_{\mu=1}^d (\varphi_{i+\textbf{e}_{ \mu }}-\varphi_{i})^2   - \frac{1}{2} m^2 \varphi_i^2 +\frac{\lambda}{4} \varphi^4_i\right] \nonumber
\label{Hphi_2}
\end{equation}
where $\textbf{e}_{ \mu } $ is the unit vector in the $ { \mu}$th direction of the $d$-dimensional lattice. 
At equilibrium and for $d\ge 2$, this model - representing a set of linearly coupled nonlinear oscillators - shows  a second order phase transition with nonzero critical temperature. This phase transition is due to a spontaneous breaking of the discrete $O(1)$, or $\mathbb{Z}_2$, symmetry.

Recently this model has been proposed as a counterexample of the topological theory \cite{KastMehta} of phase transitions. In fact, the phase transition of the $d\ge 2$ lattice $\phi^4$-model occurs at a critical value $v_c$ of the potential energy density which belongs to a broad interval of $v$-values void of critical points of the potential function. This means that the $\{ \Sigma_{v<v_c}^N\}_{v \in{\Bbb R}}$ are diffeomorphic to the  
$\{ \Sigma_{v>v_c}^N\}_{v \in{\Bbb R}}$ so that no topological change seems to correspond to the phase transition.\\
Since then, it is commonly believed that this result undermines the value of the topological theory.
However, this model undergoes a "mild" phase transition: being of the same universality class of the  Ising model, also the specific heat of the $\phi^4$-model diverges logarithmically at the transition temperature both for $d=2$ \cite{huang} and $d=3$ \cite{ising3d}, moreover, the microcanonical caloric curve (temperature versus energy) shows a markedly soft transition pattern \cite{Caiani:1998,CCCPPG} if compared to other models undergoing a continuous phase transition \cite{PettiniBook}. \\
This "mild" transition is in fact associated with a topological change of phase space and configuration space submanifolds which also is, loosely speaking, "mild": it is an asymptotic change of the number of connected components in phase and configuration spaces. It can be shown \cite{goripettini} that for any time $\tau_*\in{\Bbb R_+}$
there exists a value $N_*(\tau_*)$ of the number of degrees of freedom of the system such that for any 
$N > N_*$ 
we have $\dim H_{\tau_*}^0(\Sigma^N_E;{\Bbb R}) = 2$ for $E<E_c$ that is below the phase transition point,
whereas for $E>E_c$, that is above the phase transition point, for any $\tau\in{\Bbb R_+}$ and for any $N$ we have  $\dim H_{\tau}^0(\Sigma^N_E;{\Bbb R}) = 1$; with  $H_{x}^0(\Sigma^N_E;{\Bbb R})$ we denote the $0$-th De Rham's cohomology group of the  energy surfaces $\Sigma^N_E=H_N^{-1}(E)$ spanned by the Hamiltonian flow $\varphi^H_t$ for a time duration $t\le x$. In other words, for $E<E_c$ and $t\le x$ the energy level sets are no longer metrically transitive, that is  $\Sigma_{E,N}=\Sigma^A_{E,N}\cup\Sigma^B_{E,N}$ with $\varphi^H_t(\Sigma^A_{E,N})=\Sigma^A_{E,N}$
and $\varphi^H_t(\Sigma^B_{E,N})=\Sigma^B_{E,N}$ such that  $\Sigma^A_{E,N}\cap\Sigma^B_{E,N}=\emptyset$.\\
 The dimension of the the $0$-th De Rham's cohomology group counts the number of connected components. This means that an \textit{effective} topological change can be seen through the Hamiltonian flow on any arbitrary observational timescale, provided that $N$ is chosen accordingly. Of course, this kind of asymptotic change of topology has nothing to do  with the existence or the absence of critical points of the potential function. Moreover,  as the microcanonical measure is the invariant measure of the Hamiltonian flow, this also means that as $N$ increases the measure of the region bridging two disjoint subsets of the $\Sigma^N_E$ goes to zero. Of course the asymptotic change of topology of the energy level sets entails also the asymptotic change of topology of the potential level sets.\\
 Consequently, the hypotheses of the theorems in \cite{TH1,TH2} must be extended by assuming that the equipotential level sets $\Sigma_v^N=V_N^{-1}(v)$ beside being diffeomorphic at any finite $N$ must be also \textit{asymptotically} diffeomorphic, that is, for $N\rightarrow\infty$.\\
In so doing the $\phi^4$ model cannot be {\it logically} a counterexample of the topological theory because its phase transition actually corresponds to a major asymptotic topological change of the submanifolds $\Sigma_v^N$ of the configuration space which corresponds to an {\it asymptotic loss of diffeomorphicity} among the $\Sigma_{v<v_c}^N$ and the $\Sigma_{v>v_c}^N$ occurring in the {\it absence of critical points}.
Hence the $\phi^4$ model is no longer a counterexample of the theory and this fixes the problem.  \\
Notice that, at variance with the $MFXY$ model for which a sharp topological signature of the phase transition shows up at rather small $N$ values, the phase transition of the $d\ge 2$ lattice $\phi^4$-model is more akin to what is required by the Yang-Lee theory (that is, asymptoticity), even if tackled from the topological point of view.  
Now, taking advantage of the above mentioned results in a reverse form, if the phase space sampling through the Hamiltonian dynamics of the $\phi^4$-model is performed at a given $\widetilde{N}$ for a sufficiently long time $t \gg \tilde{\tau}(\widetilde{N})$, then also for $E<E_c$ it is 
$\dim H_t^0(\Sigma^N_E;{\Bbb R}) = 1$. In other words, this model is a good candidate for a negative check against the $MFXY$ model.

\section{Topological analysis}\label{phom}
In the following we report on  the topological analysis which begins by sampling  the configuration space of each system at different energies. Then  we apply persistent homology analysis.

\subsection{Samples of the configuration space}

We begin by  constructing samples of the configuration spaces to be studied.
For the MF$XY$ model, this is done by numerically integrating the equations of motion derived from Hamiltonian (\ref{Hmf}) with the external field set to $h=0$ for a system of $N$ spins, with $N$ up to 6000. 
The numerical integration is performed by means of a fifth-order optimal symplectic algorithm͔ \cite{McLachlan92}. 
We sampled the configuration space for the following values of the energy density $\varepsilon=E/N =0.6, 0.75, 0.88$, that is,  below, at, and above the critical energy, respectively. 
The system is initialized with a Gaussian distribution for both conjugated variables
$\left\lbrace\varphi_i,p_i\right\rbrace$.
The total angular momentum ($P=\sum_{i}p_{i}=0$) is imposed to vanish. 
Given the initial conditions for the aforementioned energies, the system dynamics is evolved for a $T=1.26 \cdot 10^7$ time steps, with an integration step of $\Delta t=0.05$. $6000$ snapshots are uniformly sampled in time after a virialization transient.
For the $ \phi^4 $ model we set $ J=1, m^2=2, d=3$ and $\lambda=0.1$ in the Hamiltonian (\ref{Hphi_2}). 
We consider a 3D cubic lattice with $8^3$ sites,  periodic boundary conditions, and an integration time step $\Delta t=0.05$. With these parameters, the use of a third order symplectic algorithm \cite{Casetti:1995} kept the relative energy fluctuations at $\Delta E/E \simeq 10^{-9}$.
Then the Hamiltonian dynamics is numerically simulated at two different values of the energy density, that is, $\varepsilon =25$ well below the transition energy density $\varepsilon_c \simeq 31$ \cite{CCCPPG}, and $\varepsilon =35$ well above $\varepsilon_c$.
\vskip -100pt

\subsection{Persistent Homology}

The main idea of persistent homology is to build  an increasing sequence of simplicial complexes, called a {\it filtration} (see \cite{Pers}), from a \textit{point cloud}, i.e. set of points embedded in a metric space. 
We report a detailed mathematical description of persistent homology in the supplementary material and refer the interested reader to \cite{Pers}. 
Here we streamline  the topological analysis.
The standard way to obtain a simplicial complex from a set of points $S$ is to construct its  $\rho$-Rips-Vietoris complex \cite{Pers}, an abstract simplicial complex that can be defined on any set of points in a given metric space ${\cal M}$. 
The $n$ simplices of the $\rho$-Rips-Vietoris complex are determined  by subsets of $n+1$ points $\lbrace p_0,\ldots, p_n \rbrace$ such that $D(p_i,\rho)\cap D(p_j,\rho)\ne \emptyset$ for all $ i\ne j \in \lbrace 0, \ldots, n\rbrace$, where $D(p,\rho)$ is ball of radius $\rho$  centered at $p$.
Persistent homology is a powerful instrument in that it does not 
select just an $\rho$ value, but rather studies how the homology of the space, and in particular of the $\rho$-Rips-Vietoris complexes, changes as $\rho$ varies.
As $\rho$ is increased, simplexes are added in the $\rho$-Rips-Vietoris simplicial complex.
A new simplicial complex is added to the filtration only when a new simplex is born along the (continuous) parameter $\rho$. 
i.e., the $\rho$-Rips-Vietoris complex has changed.
Thus the filtration  is discrete: it can be indexed by integers, useful to characterize the topological features of the space.

\subsection{Simplicial Complexes in configuration space}

In most applications of persistent homology, the parameter $\rho $ is taken to represent the Euclidean distance between points in $S$. 
In the case of physical configuration spaces we replace it by a Riemannian one.
In fact, the configuration space $M$ of a standard Hamiltonian systems (that is with quadratic kinetic energy)   equipped with the Jacobi metric \cite{PettiniBook}, is a \textit{complete Riemannian manifold}, which means that given any two points there exists a length-minimizing geodesic connecting them (Hopf-Rinow theorem).
Of course this is also the case of the mean-field  $XY$ and $\phi^4$ models, 
thus the distance among two points $P_1$ and $P_2$ in $M$ is:
\begin{equation}
d(P_1,P_2)=\int_{P_1}^{P_2} \left( [E-V(q_1,\ldots,q_N)] \sum_{k=1}^N (dq^k)^2\right )^{\frac{1}{2}}
\end{equation}
In other words, computing this distance requires solving the equations of motion with assigned initial and final conditions. In practice this is computationally very heavy. 
We therefore take advantage of the robustness of topological information with respect to metrical deformations and observe that the integral contains a non constant factor multiplying the Euclidean arc length. 
We then choose to approximate $d(P_1,P_2)$ by replacing the factor by its mean among the initial and final values:
\begin{equation}\label{distance}
d(p_1,p_2)=\frac{1}{2} ( \sqrt{E-V(p_1)} + \sqrt{E-V(p_2)} ) d_{\textit{eucl}}(p_1,p_2)
\end{equation}
\begin{equation}
d_{\textit{eucl}}(p_1,p_2)=\sqrt{\sum_{k=1}^N (q^k(p_2)-q^k(p_1))^2 }
\end{equation}

An important computational issue lies in the size of the produced simplicial complexes. 
Indeed, already for a sample of the configuration space $S$ with cardinality $N=6000$ points, the set of complexes will contain a huge number of simplices hindering efficient computation, since the number of all simplices for all dimensions up to $N-1$ scales as number of subsets of $N$, that is $2^N$.   
So, we first restrict ourselves to the study of the first two homology groups, $H_0$ and $H_1$, which allows us to consider only simplices up to dimension $2$ and then adopt a sub-sampling strategy which allows to reduce the dimension of the problem by choosing a representative subset of points $L\subset S$ without losing important topological features of the configuration space.
The sub-sampling is based on a greedy selection of landmark points called \textit{sequential maxmin} \cite{DeSilva:2005,gamble}.
In sequential maxmin, the first landmark is picked randomly from $S$. 
Inductively, if $L_{i-1} $ is the set of the first $i-1$ landmarks, then let the i-th landmark be the point of $S$ which maximizes the distance (\ref{distance}) from all the points of $L_{i-1} $.
Since the starting node is chosen at random, the resulting $L$ subsets will change if the algorithm is iterated. 
In our case, this allows us to perform a bootstrap-like procedure, by repeatedly subsampling the full point clouds and then aggregating the homological signatures detected. The results we present are obtained from 20 different sub-samples, each containing 300 points.

\section{Results}

Persistent homology computes the generators of topological features (homology groups) persisting across different scales and assigns them birth and death values related to their points of appearance and disappearance along the filtration.  
That is, when the radius $\rho$ of the balls varies, for any persistent generator $g$  (see Appendix for the formal definition) we have the value of the parameter  $\rho$ of the filtration where $g$ first appears (birth index indicated by $\beta_{g}$) and  the value where it disappears (death index indicated by $\delta_{g}$).
In this way, connected components, one-dimensional cycles, three-dimensional voids and similar higher order structures of the topological space ${\cal M}$ acquire a weight proportional to the length of their  persistence interval, $\pi_g  = \delta_g - \beta_g$.
Note that for $H_0$, $\pi_g = \delta_g$, because all (dis)connected components are already present at the beginning.  
For higher order homology groups $H_k$ the generators can instead appear and disappear freely along the filtration.

\begin{figure}
\hskip -0.6truecm\includegraphics[width=0.54\textwidth]{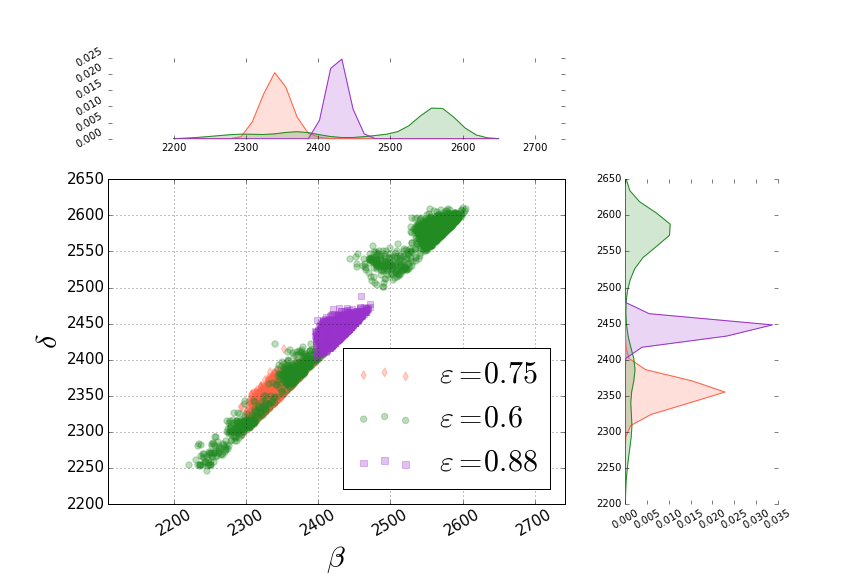}
\caption{(Color online) Persistence diagram for the $MFXY$ model. $H_1$ persistence distributions below ($\varepsilon =0.6$), at ($\varepsilon =0.75$), and above ($\varepsilon =0.88$) the phase transition.  }
   \label{subfigg}
\end{figure}

\begin{figure}
\includegraphics[width=0.45\textwidth]{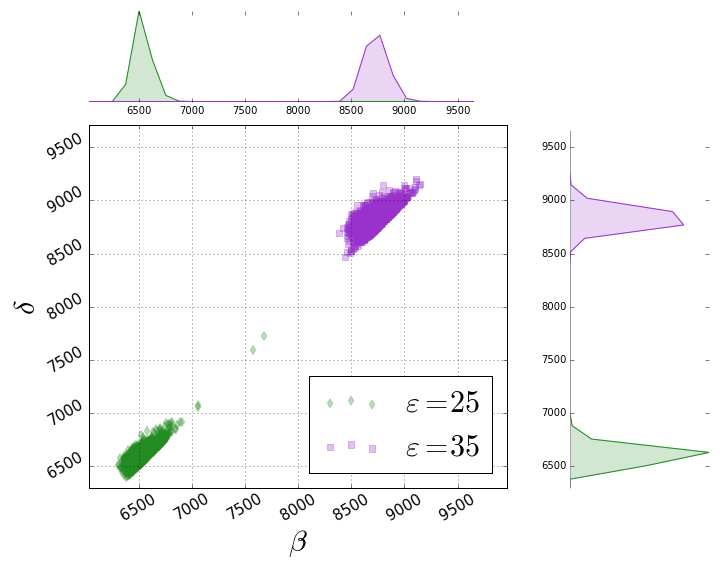}
\caption{(Color online) Persistence diagram for the $\phi^4$ model.  $H_1$ persistence distributions below ($\varepsilon =25$) and above ($\varepsilon =35$) the phase transition.  }
   \label{subfigf}
\end{figure}

In Figures \ref{subfigg} and \ref{subfigf} the basic descriptors of persistent homology, that is, persistence diagrams, are displayed for the $H_1$ generators of the $MFXY$ model and of the $\phi^4$ model, respectively. 

\begin{figure}
    \centering
\includegraphics[width=0.45\textwidth]{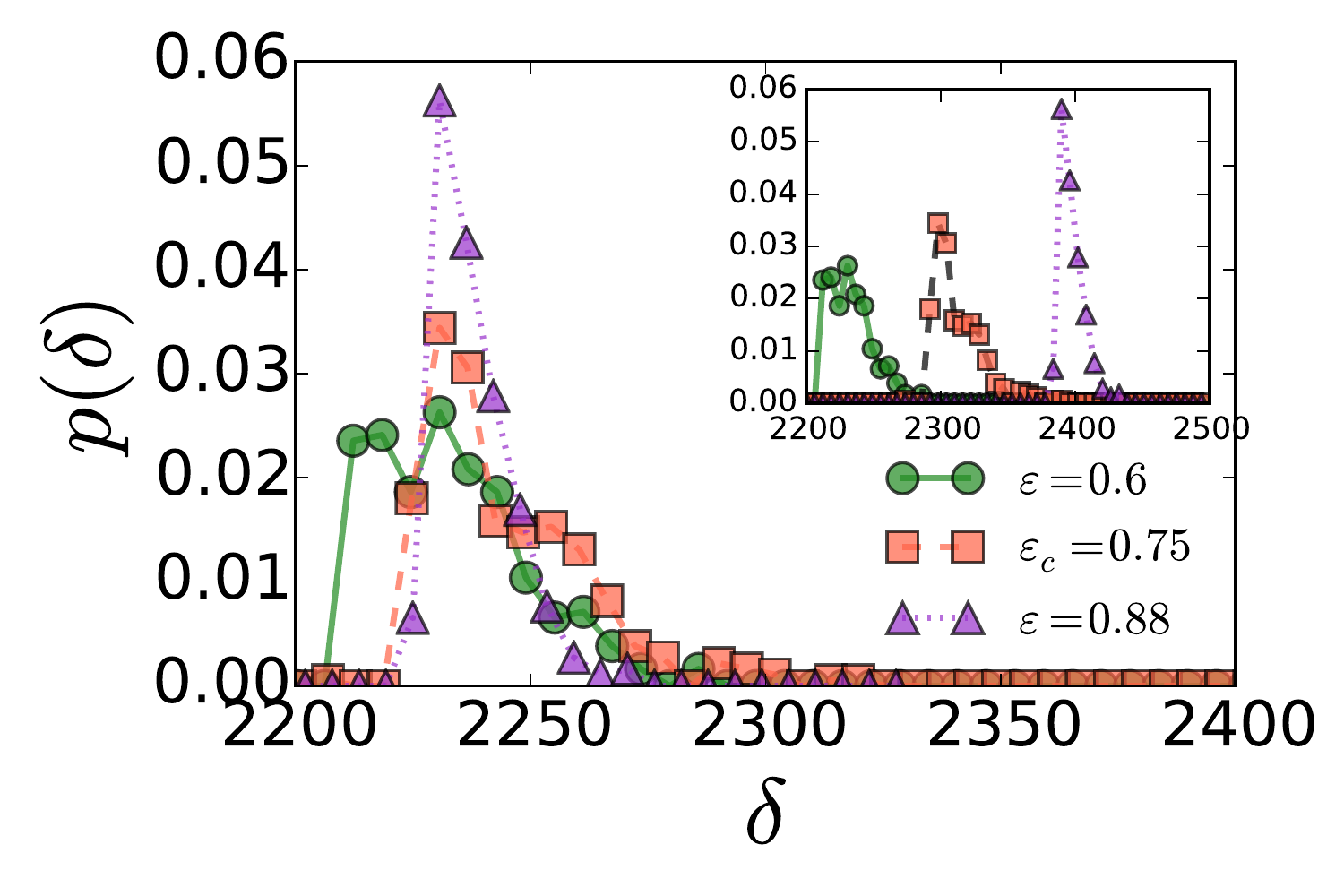}
\caption{(Color online)  Homological features of the $MFXY$ model. 
Raw (inset) and rescaled (main plot) distributions of deaths for the generators of the first homology group $H_0$.
Note that the width and shape of the distributions 
change across the transition, becoming more and more narrow as the energy is increased}
  \label{figa}
\end{figure}

\begin{figure}
\includegraphics[width=0.45\textwidth]{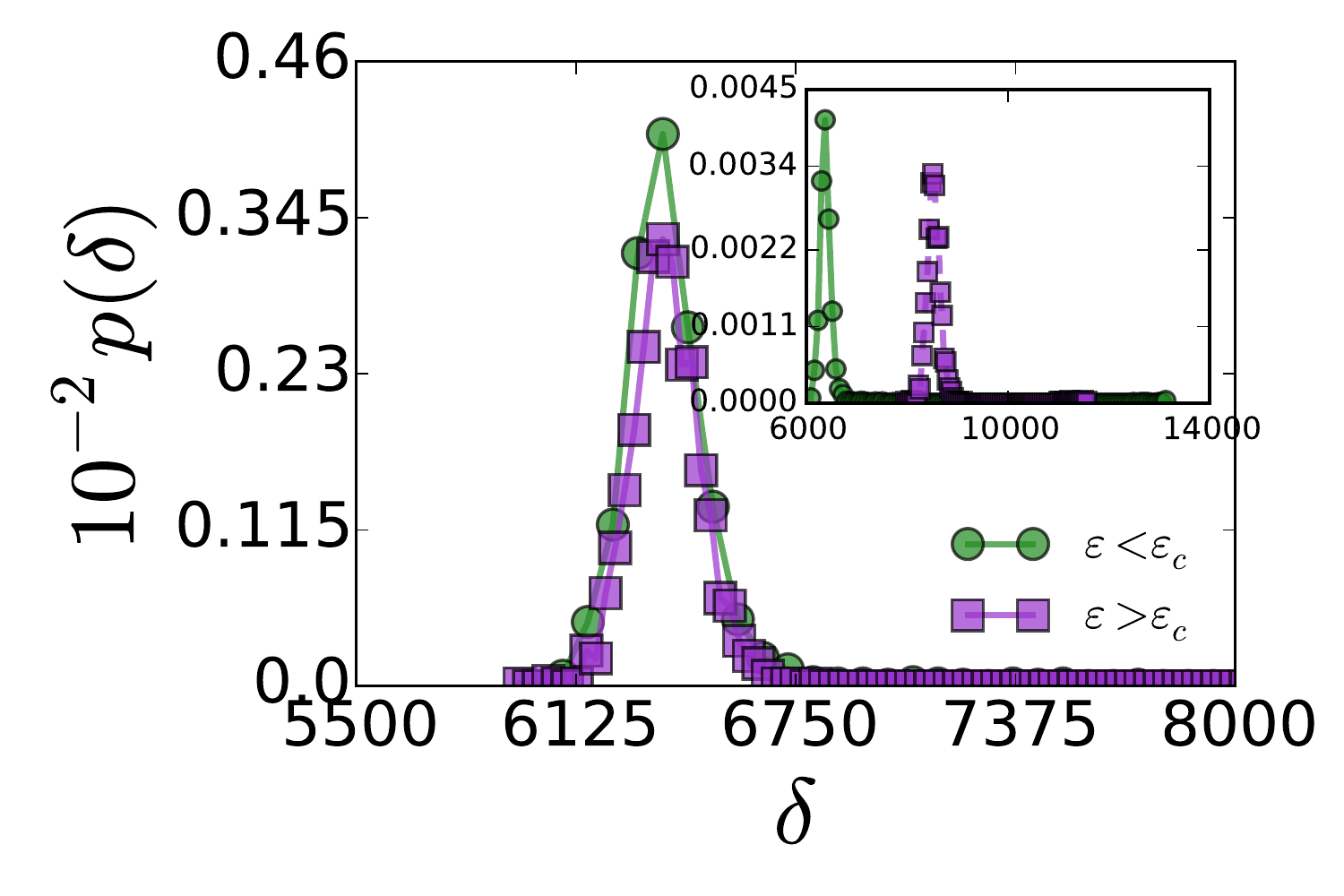}
\caption{(Color online)  Homological features of the $\phi^4$ model. 
Raw (inset) and rescaled (main plot) distributions of deaths for the generators of the first homology group $H_0$. At variance with the $MFXY$ model, here there is no appreciable change in the width and shape of the distributions across the transition. The green points refer to $\varepsilon =25 < \varepsilon_c$. The velvet points refer to $\varepsilon =35 > \varepsilon_c$.}
   \label{figb}
 \end{figure}

Usually one considers important topological features to be those associated with generators of $H_n$ such that their $\pi_g$ is large with respect to some meaningful length. \\
In our case  we do not have a given reference scale.  
We can however compare the results obtained at energies below and above the transition energy in order to look for topological signatures of a phase transition. 
We show the distributions of $\delta_g$ for the $H_0$ generators of the $MFXY$ model (Fig. \ref{figa}) and of the $\phi^4$ model (Fig. \ref{figb}).
In the former case, as the energy is increased, the peak of the distribution $\delta_g$ of $H_0$ becomes progressively narrower and centred at larger $\rho$-values. 
To the contrary, in the latter case  the peak of the distribution shifts to larger $\rho$ values at higher energies, but it does not broaden. 
In order to show that this behaviour is genuinely due to topological features and not due to the different geometrical sizes of the point clouds, we take the point cloud at the lowest energy and affinely rescale the point clouds at higher energies as to make them comparable i.e. ($d_{ij}(\varepsilon) \to \frac{\langle d(\varepsilon_0\rangle}{\langle d(\varepsilon)\rangle} d_{ij}(\varepsilon) $ where $d_{ij}(\varepsilon)$ is the distance between points $i$ and $j$ for the pointcloud at energy density $\varepsilon$.).
In this way, we can meaningfully compare the persistences of generators belonging to clouds of different size. 
Below the transition of the MFXY model, the distribution of the $H_0$ persistences of configuration space covers more scales than it does at and above the transition energy, respectively.
This broader distribution means that the corresponding point cloud is heterogeneously distributed in the embedding space $\cal{M}$ compared to the distributions, definitely more homogeneous, in the other two cases. 
No variation of the peak widths of the $H_0$ persistence distributions is observed in the case of the $\phi^4$ model.

\begin{figure}
\includegraphics[width=0.45\textwidth]{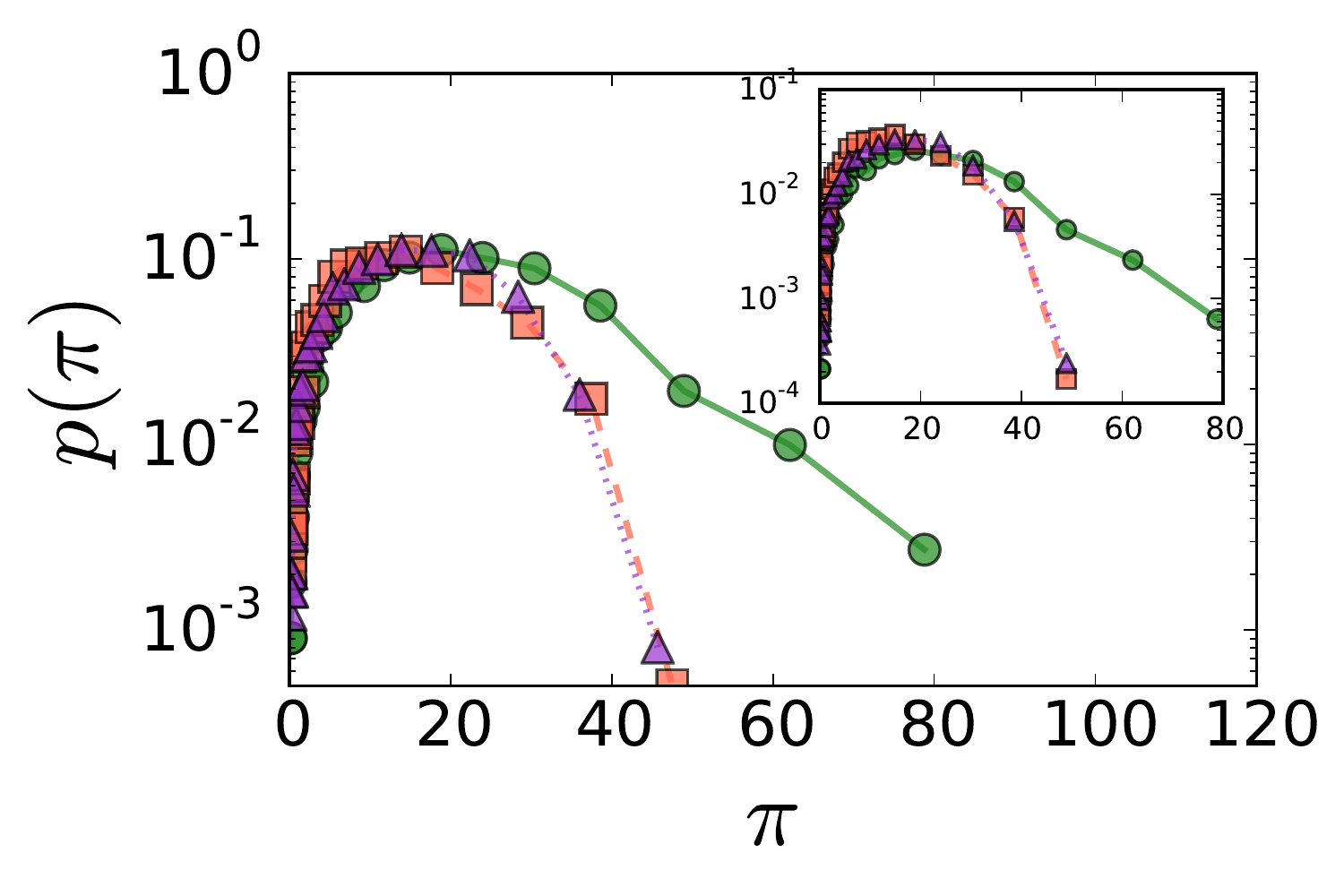}
\caption{(Color online)  Distributions of persistences for the generators of the homology group $H_1$ in the case of the $MFXY$ model. In this case the difference in functional forms for the $H_1$ persistence distribution below and above the transition is even clearer.}
   \label{subfigc}
\end{figure}

\begin{figure}
\includegraphics[width=0.45\textwidth]{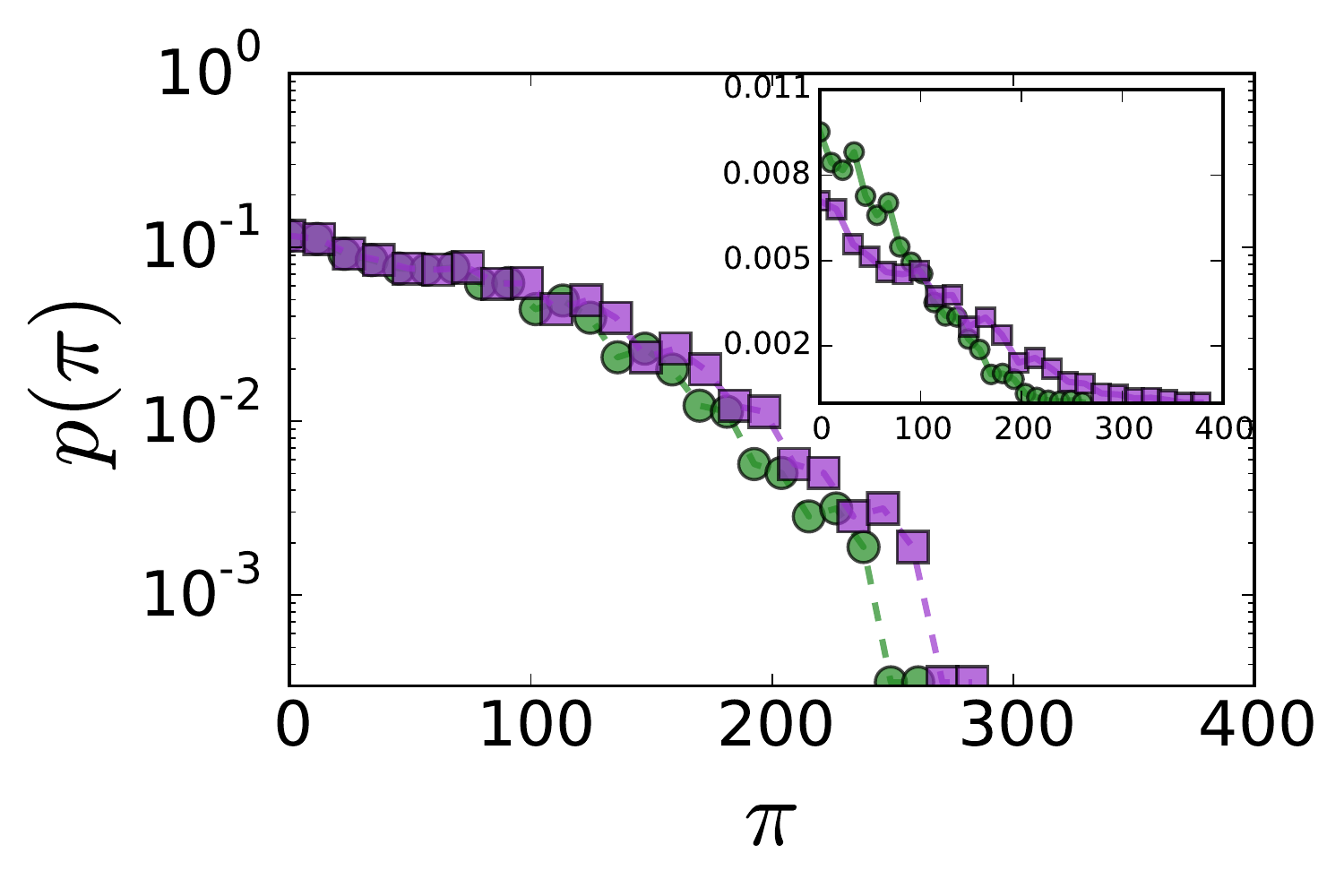}
\caption{(Color online)  Distributions of persistences for the generators of the homology group $H_1$ in the case of the $\phi^4$ model. In this case no difference is found in functional form for the $H_1$ persistence distributions below and above the transition.  }
   \label{subfigd}
\end{figure}  
 
%\caption{(Color online){\textbf{Summary of homological features of the $MFXY$ model and $\phi^4$.}  
Figures  \ref{figa} and \ref{figb} display the raw (inset) and rescaled (main plot) distributions of deaths for the generators of the first homology group $H_0$.
The rescaling is necessary to make the point clouds, sampled at different energies, comparable.
In fact, the death and birth indexes are the values of the radius of the balls where the generators appear and disappear. Thus, without the rescaling, $\beta_g$ and $\pi_g$ would reflect the size of the underlying manifold.
Note that for the $MFXY$ model the width and shape of the distributions 
change across the transition, becoming more and more narrow as the energy is increased, while there is no appreciable change in the $\phi^4$ case. 
The different topological signatures highlight the presence of a topological change in the case of the MFXY model, that is absent in the $\phi^4$ model. 
In Figures \ref{subfigc} and \ref{subfigd} the distributions of persistences for the generators of the homology group $H_1$ confirm what is found for $H_0$.
In this case the difference in functional forms for the $H_1$ persistence distribution below and above the $MFXY$ transition is even clearer, while, again, we find no differences for the $\phi^4$ model. 

Now let us comment about the hollowness { detected by} the $H_1$ homology group.
For what concerns the MFXY model, below the phase transition energy, the $H_1$ persistence distribution displays a long tail which disappears at and above the transition (Fig. \ref{subfigc}). 
We observe that the three sets of points superpose for values of $\pi$ less then approximately $25$,
this range of $\pi$ values, in the present context, can be attributed to what is commonly referred to as noise, whereas larger $\pi$-values are usually considered as bringing about meaningful topological information. Thus, the stronger persistence of meaningful cycles, which corresponds to the long tail observed below the phase transition point of the MFXY model, certainly probes a change of ``shape'' of configuration space. And this change of shape can be interpreted as the signature of a change of the dimension of high order homology groups. \\
Let us remark that the performed samplings of configuration space submanifolds are definitely sparse and they could not be other then sparse had we taken billions of points. Not to speak of the huge total number of simplexes, growing as $2^N$ with $N$ the number of sample points.
This notwithstanding, the results shown in Fig. \ref{subfigc} clearly tell us that the MFXY phase transition corresponds to a change of the topology of the configuration space submanifolds, in perfect agreement with the available theoretical knowledge.
The same concordance is found in the case of the $\phi^4$ model where we see that the difference in $H_1$ persistences disappears, in perfect agreement with a-priori known absence of topological changes of the underlying configuration space in correspondence with the phase transition.

\begin{figure}
\includegraphics[width=0.5\textwidth]{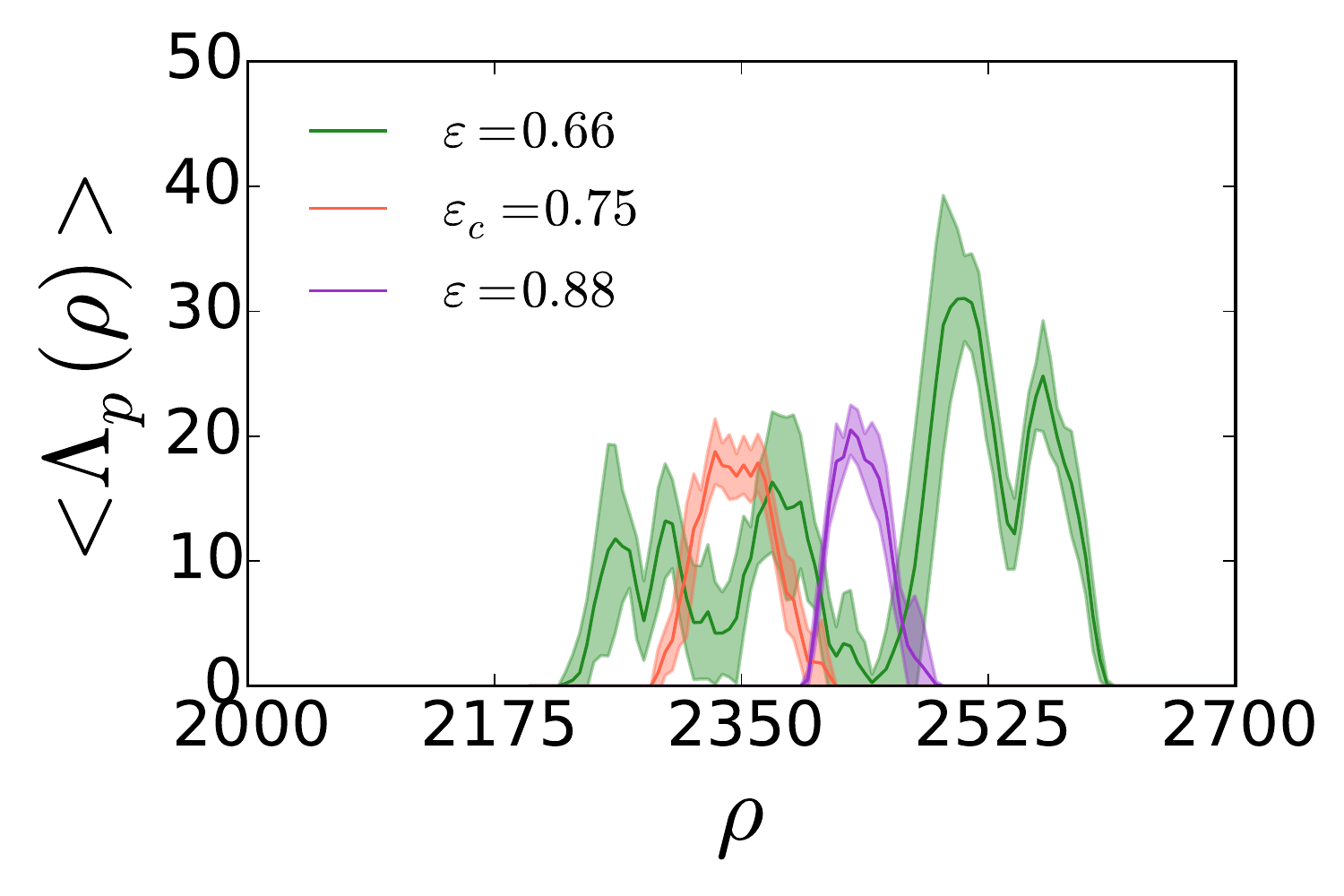}
\caption{(Color online) Average persistence landscape of the $H_1$ homology for the $MFXY$ model. $\Lambda_p$ is the average function (see text) reported  as a function of the radius $\rho$ of the balls used to construct the Rips-Vietoris simplicial complex. The "shadows" around solid lines are $95\%$ confidence band. }
   \label{persist-land}
\end{figure}

\begin{figure}
\includegraphics[width=0.5\textwidth]{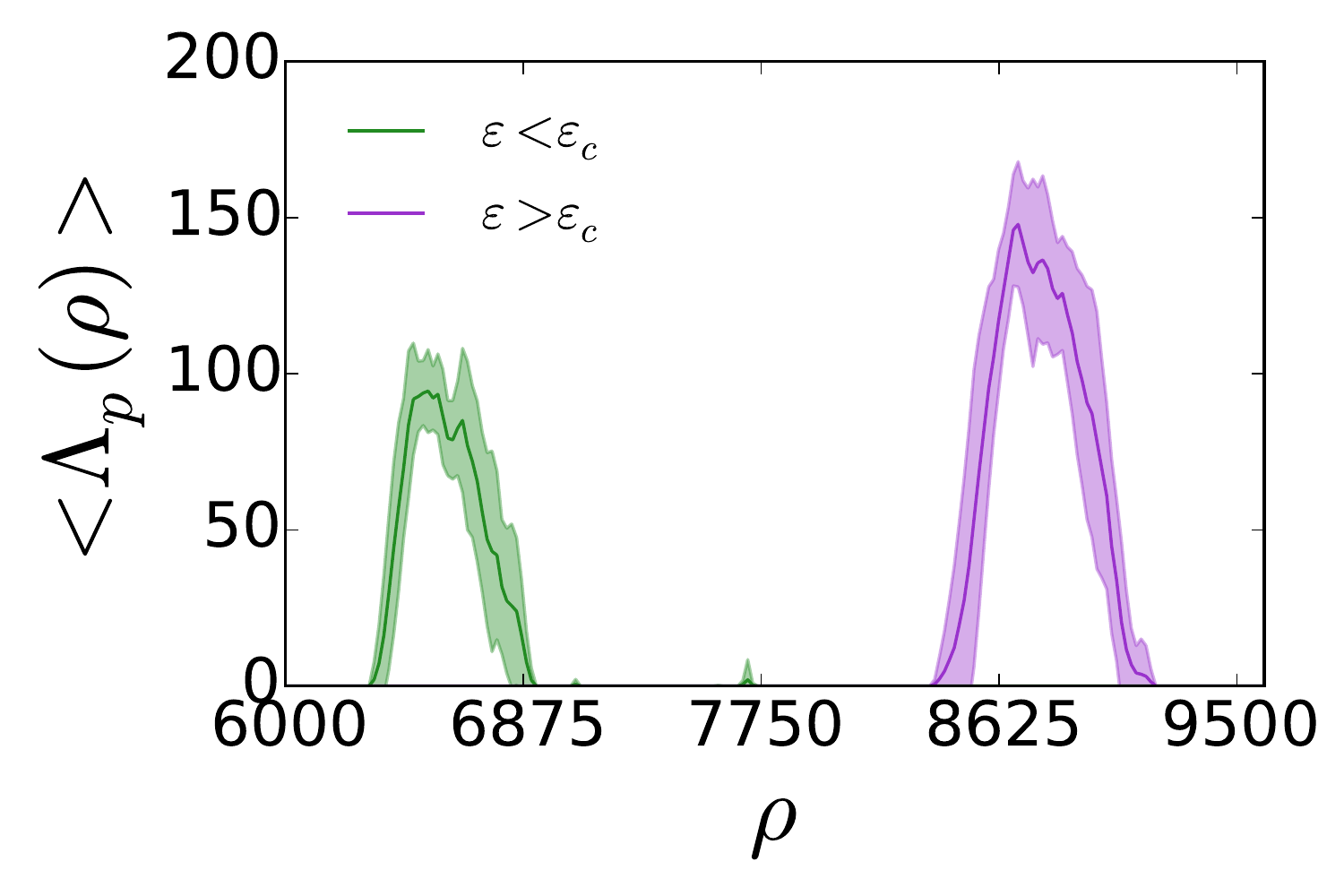}
\caption{(Color online) Average persistence landscape of the $H_1$ homology for the $\phi^4$ model. $\Lambda_p$ is the average function (see text) reported  as a function of the radius $\rho$ of the balls used to construct the Rips-Vietoris simplicial complex. The "shadows" around solid lines are $95\%$ confidence band. }
   \label{persist-land-phi4}
\end{figure}

Finally, in Figures \ref{persist-land}  and \ref{persist-land-phi4}  we show the outcomes of a different method of getting insight to the "shape" of data obtained by sampling the configuration space of the $MFXY$ and $\phi^4$ models, respectively. This is the so called \textit{persistence landscape} which combines the main tool of persistent homology method, that is, persistence diagram, with statistics \cite{bubenik}. With respect to the barcode or persistence diagram this descriptor has the technical advantage of being a function, thus allowing the use of the vector space structure of its underlying function space to apply the theory of random variables with values in this space.
Theory and details of this method can be found in Refs. \cite{bubenik} and \cite{chazal}. In practice, one proceeds by computing the $H_1$ homology for a subsample of the original dataset, then one associates to each generator a symmetric tent-shaped function peaking in the middle of the persistence interval of the corresponding generators and finally one considers the envelope of the functions defined in this way over all the generators. 
Informally, one can think of the persistence landscape as the envelope of the $\pi/4$-clockwise rotated persistence diagram (operation that can be given a proper mathematical definition) thus associating a curve $\Lambda_p(\rho)$ to each persistence diagram. 
In our case, we iterated this procedure for the different subsamples, in our case $20$ subsamples, obtaining the curve $\langle\Lambda_p(\rho)\rangle$ averaged over the samples. Each curve reported in Figure \ref{persist-land}  reports the results for different energy values: below, at, and above the phase transition point. A marked difference is again obtained above and below the phase transition in the case of the $MFXY$ model, and no relevant difference between the patterns below and above the phase transition in the case of the $\phi^4$ model, apart from a meaningless translation.  

\section{Concluding remarks}

The results reported for each model in the Figures shown in the preceding Section, and especially the comparison with those reported in Figures \ref{subfigc}, \ref{subfigd}, \ref{persist-land} and 
\ref{persist-land-phi4} are strongly supportive of the validity of the application of persistent homology to probe major topological changes in the configuration spaces of physical systems undergoing phase transitions. \\
Let us remark that the formulation of the topological theory of phase transitions stems from the combined effect of the investigation of the Hamiltonian dynamical counterpart of phase transitions on one side, and of the geometrization of Hamiltonian flows seen as geodesic flows on suitably defined Riemannian manifolds on the other side. In fact, it has been observed that the peculiar dynamical changes occurring at a phase transition correspond to special geometrical changes of the mechanical manifolds. Then it turned out that these special geometrical changes had to be due to more fundamental changes of topological kind. In other words, this theory has deep roots and rather compelling motivations \cite{PettiniBook}. Moreover, developing this unconventional viewpoint on phase transitions was of prospective interest to tackle phase transition phenomena in finite/small N systems (meso and nanoscopic systems), in the microcanonical ensemble (especially when this is not equivalent to the canonical ensemble), in the absence of order parameters (for example in gauge models, i.e. with local symmetries), in amorphous and disordered materials, in polymers and proteins, in biophysical systems, in strongly 
inhomogeneous systems. However, as mentioned in the Introduction, computational difficulties have frustrated these expectations.\\
Now the results reported in the present work show that persistent homology, by providing handy computational tools (which are presently available as open access software packages),  can lend new credit to  the prospective practical interest of the topological theory of phase transitions.
And, especially, since improvements of the numerical algorithms are continuously underway. 

Moreover, this opens many fascinating and challenging questions related with the mentioned necessarily sparse sampling of high dimensional manifolds.
It is not out of place to mention that this situation is reminiscent of Montecarlo methods which typically allow efficient estimates of multiple integrals in high dimensional spaces with very sparse samplings.
Montecarlo methods owe their efficacy to the so called \textit{importance sampling} technique, suggesting that further developments in the proposed application of the persistent homology could be found in a somewhat similar direction.

\begin{acknowledgements}
This work	was supported by the Seventh Framework Programme for Research of the European Commission under FET-Open grant TOPDRIM (Grant No. FP7-ICT-318121).
\end{acknowledgements}

\section{Appendix}

\subsection{Simplicial Complexes}
We can see a simplicial complex $X$ as a set of polyhedrons (convex hulls of linearly independent points: points, lines, triangles, tetrahedra, and higer dimensional equivalents) in ${\mathbb{R}}^N$ attached in a good way, i.e., the intersection of two polyhedrons is empty or a face of the two and all the faces of a polyhedron of $X$ is also a polyhedron of $X$.
We can also think of simplicial complexes as abstract sets, with the definition:
\begin{definition}
An (abstract) \emph{simplicial complex} is a non empty family $X$ of finite subsets, called faces, of a vertex set $V$ such that $\sigma\subset \tau \in X$ implies that $\sigma \in X.$
\end{definition}
We assume that the vertex set is finite and totally ordered. A face of $n+1$ vertices is called $n-$face, denoted by $[p_{0},\ldots, p_{n}]$, and $n$ is its dimension. We set, as usual, the dimension of the empty set as -1. 
The \textit{dimension} of a simplicial complex is the highest dimension of the faces in the complex.
\begin{figure}
\centering
\includegraphics[scale=0.23]{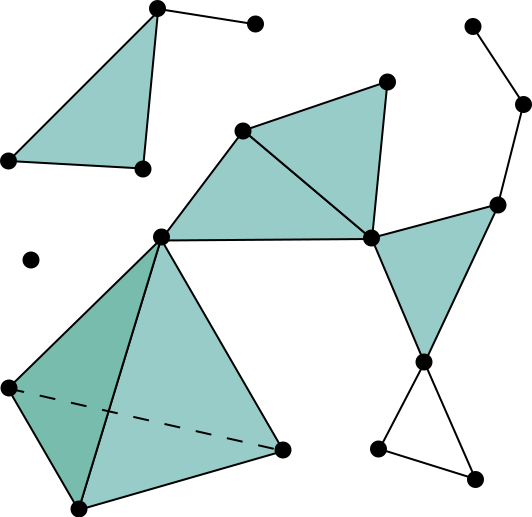}
\caption{(Color online) A graphic representation of a simplicial complex}\label{fig:1}
\end{figure}

\subsection{Simplicial Homology}
Let us fix a field $k$. In the following, by vector space we intend $k-$vector space. 
Given a simplicial complex $X$ of dimension $d$, for any $n$ such that $0\leq n \leq d$ consider the vector space $\CC_n:=\CC_n(X)$ of all the linear combinations of $n$-faces of $X$ with coefficients in $k$. Elements in $\CC_n$ are called $n$\textit{-chains}.\\

The \textit{boundary} operators are the linear maps sending a $n$-face to the alternate sum of its $(n-1)$-faces, i.e.,
\begin{eqnarray*}
\partial_n: \CC_n & \longrightarrow & \CC_{n-1}\\
\text{[}p_{0},\ldots, p_{n}\text{]} & \rightarrow & \sum_{j=0}^n(-1)^i\text{[}p_{0}, \ldots, p_{{j-1}}, p_{{j+1}},\ldots, p_{n}\text{]}.\\
\end{eqnarray*}
They share the property $\partial_{n-1}\circ\partial_n=0$. The subspace $\ker\partial_n$ of $\CC_n$ is called the vector space of $n$-\textit{cycles} and denoted by $\ZZ_n:=\ZZ_n(X)$, with by convention $\ZZ_0 = \emptyset$.
The subspace  $\mathrm{Im}\,\partial_{n+1}$ of $\CC_n$, is called the vector space of $n$-\textit{boundaries} and denoted by $\BB_n:=\BB_n(X)$, with by convention $\BB_d = \emptyset$. The property $\partial_{n-1}\circ\partial_n=0$ is then equivalent to $\BB_n\subseteq \ZZ_n$ for all $n$.
\begin{definition}
For $0 \leqslant n \leqslant d$, the $n-$th simplicial homology space of $X$, with coefficients in $k$, is the vector space $\HH_n:=\HH_n(X):=\ZZ_n/\BB_n$.
We denote by $\beta_n:=\beta_n(X)$ the dimension of $\HH_n$ which is usually called the $n$-th Betti number of $X$. 
\end{definition}
Let us see two examples.
First, let us consider the simplicial complex $X$ consisting of a triangle $[p_{1}p_{2}p_{3}]$ and all its edges and vertices (i.e., $X=\{[p_1p_2p_3],[p_1p_2],[p_1p_3],[p_2p_3],[p_1],[p_2],[p_3]\}$). The boundary of the 2-simplex $[p_{1}p_{2}p_{3}]$ is
$$\partial_2 ([p_{1}p_{2}p_{3}])=[p_{2}p_{3}]-[p_{1}p_{3}]+[p_{1}p_{2}]$$
that is a one-chain whose boundary is
\begin{eqnarray*}
\partial_1 ( &[p_{2}p_{3}]-[p_{1}p_{3}]+[p_{1}p_{2}]) =[p_{3}]-[p_{2}]+\\
&+[p_{1}]-[p_{3}]+[p_{2}]-[p_{1}] =0.
\end{eqnarray*}
Therefore $Z_1=B_1$ is the vector space generated by $[p_{2}p_{3}]-[p_{1}p_{3}]+[p_{1}p_{2}]$%, $Z_2=\emptyset$, $B_0$ is the vector space generated by $(p_1-p_2)$ and $(p_2-p_3)$
, so $H_1=\emptyset$ and $\beta_1=0$.

After let us consider the simplicial complex $X'$ consisting of all the edges and vertices of the triangle but without the face $[p_{1}p_{2}p_{3}]$ (i.e., $X' = X/[p_1p_2p_3]$). Therefore $Z'_1$ is generated by $[p_{2}p_{3}]-[p_{1}p_{3}]+[p_{1}p_{2}]$ whereas $B'_1=\emptyset$. So $H'_1=Z'_1$ and $\beta'_1=1$.
Comparing the two examples, we see that by eliminating the two-face from $X$ (roughly speaking, punching hole in the triangle) a generator of $H_1$ is created. In conclusion, the homology spaces characterize the presence of holes in simplicial complexes.
Indeed, the $0$-th Betti number is the number of connected components of $X$, the first Betti number is the number of generators of two dimensional (poligonal) holes, the third Betti number is the number of generator of three dimensional holes (convex polyhedron), etc.

\subsection{Persistent Homology}
The starting point in persistent homology is a filtration.
As in \cite{Pers}, we call a simplicial complex $X$ filtered if we are given a family of subspaces
$\{X_v\}$ parametrized by $\N$, such that $X_v\subseteq X_w$ whenever $v\leq w$ {\color{violet} and $X_v$ is a simplicial complex}. 
The family $\{X_v\}$ is called a \textit{filtration}. 

There are many ways to construct a filtration from a point cloud or a network. The most popular filtration for data analysis is the \textit{Rips-Vietoris filtration} \cite{Pers}. 

The Rips-Vietoris complex is a simplicial complex associated to a set of points in a metric space in the following way: every point $p$ is the center of a radius $\rho$ ball $D(p,\rho)$ and $n+1$ points $\{p_0,\ldots, p_n\}$ determine a $n-$face in the Rips-Vietoris complex if the corresponding radius $\rho$ balls intersect two by two, i.e $D(p_i,\rho)\cap D(p_j,\rho)\ne \emptyset$ for all $ i\ne j \in \{0\ldots n\}$.
Clearly the Rips-Vietoris complex depends on the parameter $\rho$ and if $\rho_1<\rho_2$ the complex with $\rho_1$ radius balls is contained in the complex with $\rho_2$ radius balls.
To the growth of $\rho$ we obtain an increasing sequence of simplicial complexes, a filtration, the Rips-Vietoris filtration. In this context persistent topological features of the filtration are considered as features of the point cloud. 

\begin{figure}
\centering
\includegraphics[scale=0.22]{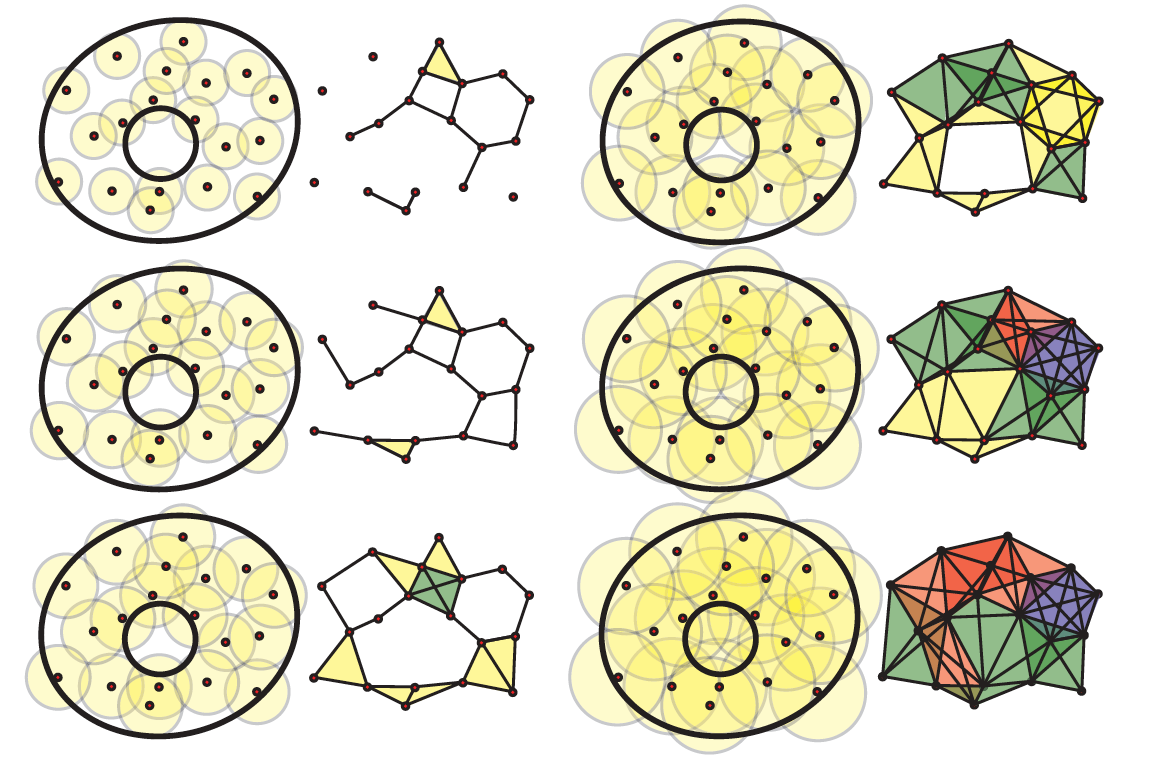}
\caption{(Color online) Rips Complex Filtration.  Reproduced with owner's permission from: Ghrist, Robert. "Barcodes: the persistent topology of data." Bulletin of the American Mathematical Society 45.1 (2008): 61-75.}\label{fig:2}
\end{figure}

The following basic properties of the algebraic structure of persistent homology hold:
\begin{proposition}
Let $X$ and $Y$ be two simplicial complexes, a simplicial map $f: X \rightarrow Y$ 
is a map sending vertices of $X$ to vertices of $Y$ and faces of $X$ to faces of $Y$. Then $f$
determines a linear map between the homology groups $H_i (f): H_i(X)\rightarrow H_i(Y)$ for all $i$.
\end{proposition}
From which it makes sense the following.
\begin{definition}
The \textit{persistent homology module} of a filtration is given by the direct sum of the homology groups of the simplicial complexes $H_n(X_v)$ and the linear maps $i_{v,w}: H_n(X_v)\rightarrow H_n(X_{w})$ induced in homology by the inclusions $X_v\hookrightarrow X_w$ for all $v\leq w$. 
\end{definition}

Following \cite{Pers}, this system is called a module because the direct sum of vector spaces $H_n=\oplus_v H_n(X_v)$ has a $k[x]-$module structure via an algebraic action given by $x\cdot m:=i_{v,v+1}(m)$ for $m\in H_n(X_v)$. The linear maps $i_{v,v+1}$ are not always injective.
A persistent homology generator is a generator of $H_n$ as $k[x]-$module, i.e an element $g \in H_n(X_v)$ such that there is no $h \in H_n(X_w)$ for $ w<v$ with the property that $x^{v-w}h=g.$
By the structure theorem on modules over principal ideal domains, the isomorphism class of a $k[x]-$module is completely determined by the degree of each generator $g$ (birth of the generator $\beta_{g}$) and the degree in which the generator is annihilated by the module action (death of the generator $\delta_{g}$). The persistence (lifetime) of a generator is measured by $p_g:=\delta_g-\beta_g$.

Persistent homology modules can be computed using libraries like {\it javaPlex} (Java) or {\it Dionysus} (C++), which are both available from the Stanford's CompTop group website (\url{http://comptop.stanford.edu/}).

\hfill
%\newpage
 %%%%%%%%%%%%%%%%%%%%%%%%%%%%%%%%%%%%%%%%%%%%%%%%%%%%%%%%%%%%%%%%%%%%%%
 % BIBLIOGRAPHY
\bibliographystyle{unsrt}

\bibliography{phase_bib}

\end{document}